\documentclass[12pt]{JHEP3}
\usepackage{amssymb,amsmath,amsthm}
\usepackage{mathrsfs} 
\usepackage{epsfig,multicol,bbm}
\def\TODAY{14 January 2011; 20 January 2011}

\title{Quasi-normal frequencies: \\ Key analytic results}
\author{Petarpa Boonserm\\
Department of Mathematics, Faculty of Science,
Chulalongkorn University, Phayathai Road, Pathumwan
Bangkok 10330, Thailand\\
E-mail: \email{petarpa.boonserm@gmail.com}
}
\author{Matt Visser\\
School of Mathematics, Statistics, and Operations Research, \\
Victoria University of Wellington, 
Wellington, New Zealand\\
E-mail: \email{matt.visser@msor.vuw.ac.nz}
}
\received{\TODAY;  \LaTeX-ed \today}                  
\abstract{ The study of exact quasi-normal modes [QNMs], and their associated quasi-normal frequencies [QNFs], has had a long and convoluted history --- replete with many rediscoveries of previously known results. In this article we shall collect and survey a number of known analytic results, and develop several new analytic results --- specifically we shall provide several new QNF results and estimates, in a form amenable for comparison with the extant literature. Apart from their intrinsic interest, these exact and approximate results serve as a backdrop and a consistency check on ongoing efforts to find general model-independent estimates for QNFs, and general model-independent bounds on transmission probabilities. Our calculations also provide yet another physics application of the Lambert $W$ function.  These ideas have relevance to fields as diverse as black hole physics, (where they are related to  the damped oscillations of astrophysical black holes, to greybody factors for the Hawking radiation, and to more speculative state-counting models for the Bekenstein entropy), to quantum field theory (where they are related to Casimir energies in unbounded systems),  through to condensed matter physics, (where one may literally be interested in an electron tunelling through a physical barrier). 
}

\keywords{quasi-normal modes, QNMs, quasi-normal frequencies, QNFs,  transmission amplitudes, reflection amplitudes, transmission probabilities, reflection probabilities, tunnelling, Eckart potential, P\"oschl--Teller potential, Rosen--Morse potential, Morse potential, Manning--Rosen potential, Hulthen potential, Tietz potential, Hua potential.
\\[10pt]
PACS:  04.70.-s; 03.65.-w; 03.65.Ge; 02.30.Gp}

\preprint{}
\begin{document}

\def\d{{\mathrm{d}}}
\def\sech{{\mathrm{sech}}}
\def\Re{{\mathrm{Re}}}
\def\Im{{\mathrm{Im}}}
\clearpage

\section{Introduction}

The investigation of  exact quasi-normal modes [QNMs], and their associated quasi-normal frequencies [QNFs],  has had a long and convoluted history, replete with many rediscoveries of previously known results~\cite{Boonserm}. Physically there are strong connections to black hole physics, at least some versions of quantum gravity, and quantum field theory. There are also strong connections between QNMs, QNFs, and transmission resonances --- which is why analytic understanding of QNFs is intimately related to knowledge of exact solutions for transmission  amplitudes and transmission probabilities. In this article (focussing mainly on QNFs) we shall collect and survey a number of known analytic results,  develop several new analytic results,  and develop several new perturbative estimates, in a form amenable to comparison with the extant literature.

In particular we shall first discuss standard textbook fare such as the delta-function potential, double-delta-function potential,  and asymmetric double-delta-function potential~\cite{Baym, Gasiorowicz, Galindo, Landau, Schiff, Messiah, Morse-Feshbach, Merzbacher, Brandsen, Liboff, Capri, Shankar, Birrell-Davies}, using them to illustrate transmission resonances, damped modes, and QNFs --- with already some significant new results even at this elementary level. We shall develop exact analytic results for the QNFs of the double-delta-function potential in terms of the Lambert $W$ function, and develop some  perturbative estimates for the QNFs of the asymmetric double-delta-function potential. 
Secondly, we shall turn to  the step barrier and  the symmetric and asymmetric rectangular potential barriers~\cite{Baym, Gasiorowicz, Galindo, Landau, Schiff, Messiah, Morse-Feshbach, Merzbacher, Brandsen, Liboff, Capri, Shankar, Birrell-Davies}, again extracting some new results on QNFs and damped modes.

Finally we shall then turn to something more challenging,   the Eckart potential~\cite{Eckart} and its common simplifications --- the  $\mathrm{tanh}(\cdot)$  and $\mathrm{sech}^2(\cdot)$ potentials.  The history of the Eckart potential is particularly complicated: Apart from the special case of the Morse potential (which pre-dates Eckart's work by one year~\cite{Morse}), other special cases and equivalent reformulations of Eckart's results (such as the Rosen--Morse~\cite{Rosen-Morse}, P\"oschl--Teller~\cite{Poeschl-Teller}, Manning--Rosen~\cite{Manning-Rosen}, Manning~\cite{Manning}, Hulthen~\cite{Hulthen}, Tietz~\cite{Tietz}, and Hua~\cite{Hua} potentials) post-date Eckart's work by anything ranging from several years to six decades. 
While the QNFs for the  $\mathrm{sech}^2(\cdot)$ potential are quite standard and well-known, the QNFs for the $\mathrm{tanh}(\cdot)$  potential and general Eckart potential are at best only implicitly given in the extant literature. 

Apart from their intrinsic interest, these exact results serve as a backdrop and a consistency check on ongoing efforts to find general model-independent bounds on transmission probabilities~\cite{bounds1, bounds2, greybody, Miller-Good, analytic-bounds, Shabat-Zakharov}, and general model independent estimates for QNFs~\cite{Medved1, Medved2, Padmanabhan1, Padmanabhan2, Jozef1, Jozef2, Jozef3, Jozef4}. Thus these ideas have relevance to fields as diverse as black hole physics (where they are related to the QNFs controlling the damped oscillations of astrophysical black holes, to greybody factors for the Hawking radiation, and more speculatively to state-counting models for the Bekenstein entropy), quantum field theory (where analysis of the QNFs is an important technical step in calculating Casimir energies for unbounded systems~\cite{Nesterenko1, Nesterenko2}), through to condensed matter physics (where one may literally be interested in an electron tunelling through a physical barrier).

We will be particularly interested in the QNFs (complex energies corresponding to purely outgoing waves in both spatial directions)~\cite{Jozef1, Jozef2, Jozef3, Jozef4, Nesterenko1, Nesterenko2, Kokkotas, Nollert, Berti, Dreyer, Natario, Das1, Das2, Beyer}, and the closely related transmission resonances (real energies where the transmission probability is unity).
Formally,  QNFs are most easily found by looking for complex frequencies where the transmission \emph{amplitude} becomes infinite.

Experience obtained many independent fields (including black hole physics)  has shown that it is quite common for the QNFs to be approximately of the form
\begin{equation}
\omega_n =  \hbox{(offset)}  +  i n \hbox{(gap)} + \mathcal{O}(1/n),
\end{equation}
where the ``offset'' is generally a complex number and the ``gap'' is typically a real number~\cite{Medved1, Medved2, Padmanabhan1, Padmanabhan2, Jozef1, Jozef2, Jozef3, Jozef4, Nesterenko1, Nesterenko2, Kokkotas, Nollert, Berti, Dreyer, Natario, Das1, Das2}. We shall be particularly interested in checking for such asymptotic behaviour in the specific models we consider below. In particular the asymptotic behaviour of the QNFs for the double-delta potential is more subtle than one might have expected.

\section{Conventions}

To set the stage, we are interested in solving the time--independent Schr\"odinger equation 
\begin{equation}
\bigg[-{\hbar^2 \over 2m} \, {\d^2 \over \d x^2} + V(x) \, \bigg] \, \psi(x) = E \, \psi(x).
\end{equation}
In regions where the potential is zero the wavefunction takes the form
\begin{equation}
\psi(x) =  {\exp(\pm i k x )\over\sqrt{k}};  \qquad E = {\hbar^2 k^2\over 2m}.
\end{equation}
We shall define outgoing modes by
\begin{equation}
\psi(x) \to   {\exp( - i k |x|   )\over\sqrt{k}};  \qquad |x|\to\infty.
\end{equation}
QNFs are in this non-relativistic context more properly called quasi-normal wavenumbers, and correspond to $Im(k)\geq 0$ so that
\begin{equation}
|\psi(x)| \to   {\exp( Im(k) |x|   )\over\sqrt{k}} \to \infty;  \qquad |x|\to\infty.
\end{equation}
(Other sign and phase conventions are also in common use, and there is no universal agreement as to the ``best'' set of conventions,  but all authors agree that the quasi-normal modes are growing at spatial infinity and so are non-normalizable.)
Once one has found the quasi-normal wavenumbers,  the associated quasi-normal energies are (in this non-relativistic context) simply given by 
\begin{equation}
E_\mathrm{QNF} = \hbar^2 k_\mathrm{QNF}^2 /(2m).
\end{equation}
For symmetric situations where the potential has the same limit $V(x)\to V_\infty$ at $x\to\pm\infty$ the scattering problem is characterized by the asymptotic behaviour
\begin{equation}
 \psi(x) \rightarrow \left\{ 
 \begin{array}{ll}
\displaystyle  t  \; {\exp( -ik_\infty x   )\over\sqrt{k_\infty}} &\quad [x\to+\infty];
 \\
 \\
\displaystyle  {\exp( -ik_\infty x   )\over\sqrt{k_\infty}}  + r \; {\exp( ik_\infty x   )\over\sqrt{k_\infty}} &  \quad [x\to-\infty].
 \end{array} \right.
\end{equation}
Should the potential have distinct limits $V(x) \to V_{\pm\infty}$ as  $x\to\pm\infty$ then one needs to distinguish the asymptotic wavenumbers $k_{\pm\infty}$ and the the scattering problem is characterized by the more complicated asymptotic behaviour
\begin{eqnarray}
\psi(x) \rightarrow \left\{
\begin{array}{ll}
\displaystyle t  \; {\exp( -ik_{+\infty} x   )\over\sqrt{k_{+\infty} }} &\quad [x\to+\infty];
\\
\\
\displaystyle {\exp( -ik_{-\infty} x   )\over\sqrt{k_{-\infty} }}  + r \; {\exp( ik_{-\infty} x   )\over\sqrt{k_{-\infty} }}   & \quad [x\to-\infty].
\nonumber\\
\end{array} \right.
\\
\end{eqnarray}
If one wishes to work directly in a relativistic context then instead of the Schr\"odinger equation  the relevant PDE is the closely related
\begin{equation}
 \bigg[+{\d ^2 \over\d t^2} -  {\d^2 \over \d x^2} + V(x) \, \bigg] \, \psi(t,x) =0.
\end{equation}
For relativistic wavefunctions of the form $\psi(t,x)=\exp(i\omega t) \psi(x)$ this reduces to 
\begin{equation}
 \bigg[- {\d^2 \over \d x^2} + V(x) \, \bigg] \, \psi(x) = \omega^2 \, \psi(x).
\end{equation}
This is formally equivalent to the situation for the non-relativistic time-independent Schr\"odinger equation under the substitutions $\hbar^2/(2m)\to 1$ and  $E\to\omega^2$. Thus there is no particular need to treat the relativistic situation separately.  We shall phrase the discussion below in terms of the non-relativistic problem with the understanding that the fully relativistic situation can easily be recovered with the appropriate substitutions.


\section{Selected potentials leading to exact results}

We start by looking at some extensions of textbook results and connecting them back to the commonly conjectured ``$\hbox{(offset)} + i n \hbox{(gap)}$'' behaviour for highly damped QNFs. 

\subsection{Delta-function potential}

For a delta function potential take
\begin{equation}
 V(x) = \alpha \, \delta(x).
\end{equation}
Since the potential is zero for $x\neq 0$ we can in this region relate the energy to the wavenumber via
\begin{equation}
E = {\hbar^2 \; k^2\over 2m}.
\end{equation}
It is extremely useful to define
\begin{equation}
k_{0} = {m \alpha \over \hbar^2}.
\end{equation}
In this case the transmission amplitude is well known to be (see for instance Baym~\cite{Baym} or Gasiorowicz~\cite{Gasiorowicz})
\begin{equation}
t = {1 \over 1 - \displaystyle{ik_0\over k}},
\end{equation}
where this expression holds even for complex $k$.  For  real $k$ (and hence real $E$)  the transmission probability is (see for instance~\cite{Baym, Gasiorowicz})
\begin{equation}
T = {1 \over 1 +\displaystyle {k_0^2 \over k^2}}.
\end{equation}
There are no transmission resonances for this potential. (Speaking rather loosely one could formally view the limit $k\to\infty$ as a transmission resonance, as $T\to1$ in this limit --- but this is not standard nomenclature.)

The QNFs are located by finding the poles in the transmission amplitude $t$. There is only  one QNF, which is pure imaginary, (and so strictly speaking should be called a damped mode rather than a QNF). It is given by 
\begin{equation}
 k_\mathrm{QNF} = i k_0;  \qquad E_\mathrm{QNF} = - {\hbar^2 k_0^2\over2m}.
\end{equation}
For $\alpha>0$ the delta function is repulsive, and $Im(k)>0$ for the formal QNF, corresponding to a damped-mode QNF. On the other hand, for $\alpha<0$ the delta function is attractive --- then $Im(k)<0$ for the formal QNF, so the formal QNF is actually a bound state. 
Because the width of the delta function is zero,  the ``gap'' is infinite, and the ``higher'' QNFs are effectively driven off to imaginary infinity --- only the ``lowest'' QNF survives. 


\subsection{Double-delta-function potential}
For the double delta function
\begin{equation}
V(x) = \alpha \{\delta (x - a) + \delta(x + a)\} ,
\end{equation}
the transmission amplitude is known to be (see for instance Galindo and Pascual~\cite{Galindo})
\begin{equation}
t = {k^2 \over (k - i k_{0})^2 + k_{0}^2 \, \exp(-4 i k a)},
\end{equation}
even for complex $k$. Here we again set
\begin{equation}
k_{0} = {m \alpha \over \hbar^2}.
\end{equation}
For real $k$ the transmission probability is (see for instance~\cite{Galindo})
\begin{equation}
T = {1 \over 1 + 4 \left[\displaystyle{k_0 \over k} \, \cos (2k a) + \displaystyle{\bigg({k_0 \over k}\bigg)^2 }\sin (2k a) \right]^2}.
\end{equation}
Equivalently
\begin{equation}
T = {1 \over 1 + 4\;  \displaystyle {k_0^2\over k^4} \; \left[ k\, \cos (2k a) + k_0 \, \sin (2k a) \right]^2}.
\end{equation}
The transmission resonances are located by
\begin{equation}
T=1  \quad \Longleftrightarrow \quad k = - k_0 \tan (2ka).
\end{equation}
This is a transcendental equation with an infinite family of exact solutions. For large real $k$ approximate solutions are
\begin{equation}
2 k a \approx \left(n+{1\over2}\right) \pi.\vphantom{\Bigg|}
\end{equation}
The QNFs are located by 
\begin{equation}
 t = \infty \quad \Longleftrightarrow \quad  (k - i k_{0})^2 + k^2_0 \exp (-4 ika) = 0,
\end{equation}
 so that
 \begin{equation}
 \exp(4 ika) =  \bigg(1 + {i k \over k_0}\bigg)^2.
 \end{equation}
 Alternatively
 \begin{equation}
 \exp(-2 ika) =  \pm \bigg(1 + {i k \over k_0}\bigg).
 \end{equation}
One obvious formal solution is $k=0$, but this is physically uninteresting and does not correspond to a true physical QNF. The general solution to the QNF condition is
\begin{equation}
k_\mathrm{QNF} = i \left\{ k_0 -  {W( \pm 2k_0a \; e^{2k_0 a} )\over 2 a} \right\}.
\end{equation}
Here $W(x)$ is the Lambert $W$ function implicitly defined by $W(x) \, e^{W(x)} = x$. (For a general discussion of the Lambert $W$ function see~\cite{Corless-et-al, Corless2, Corless3, Kheyfits, Veberic, nist}.)  Now the Lambert $W$ function is, like the logarithm or many other complex functions, a multi-valued function depending on which particular branch one is dealing with.  The real branches are $W_0(x)$ for $x\in(-e^{-1},\infty)$ and $W_{-1}(x)$ for $x\in(-e^{-1},0)$. 
For these two branches $W_{\{0,-1\}}(x e^x) = x$, while on the other hand $W_{\{0,-1\}}(-x e^x)$ is nontrivial.  This now leads to one trivial QNF at $k=0$, plus a second nontrivial QNF at 
\begin{equation}
k_\mathrm{QNF} = i \left\{ k_0 -  {W_{\{0,-1\}}(- 2k_0a \; e^{2k_0 a} )\over 2 a} \right\}.
\end{equation}
This particular QNF is pure imaginary (either a damped mode or a bound state) whenever the $W$ function is real, that is whenever 
\begin{equation}
2 a k_0 < W(e^{-1}) = 0.2785...
\end{equation}
This particular exact result for $k_\mathrm{QNF}$ is already a rather nontrivial analytic result that does not appear to be well-known.  As $a\to0$, so that the 2 delta functions merge and $W(x) = x + \mathcal{O}(x^2)$, this reproduces the previous single delta function result. 
In addition, from the other (guaranteed complex) branches of the $W$ function one obtains an infinite tower of QNFs --- specifically
\begin{equation}
k_\mathrm{QNF}(n) = i \left\{ k_0 -  {W_n( \pm 2k_0a \; e^{2k_0 a} )\over 2 a} \right\}.
\end{equation}
The corresponding quasi-normal energies are
\begin{equation}
E_\mathrm{QNF}(n) = - {\hbar^2\over2m} \left\{ k_0 -  {W_n( \pm 2k_0a \; e^{2k_0 a} )\over 2 a} \right\}^2.
\end{equation}
Thus already in this simple situation, as soon as one has non-zero separation $a$ between the two delta-functions, then an infinite tower of QNFs  arises. As the ``width'' $a$ shrinks to zero all but the lowest lying QNF are driven off to infinity.

Quantitative asymptotic information can be extracted by first noting 
\begin{equation}
W_n(x) = - \ln W_n(x) + \ln x + i 2\pi n.
\end{equation}
By recursively iterating this expression one can derive the Comtet expansion (see for example~\cite{Corless2}). The leading terms of this Comtet expansion can best be recast as
\begin{equation}
W_n(x) = \ln x + i 2\pi n - \ln(\ln x + i 2\pi n) + \mathcal{O}\left\{ {\ln(\ln x + i 2\pi n)\over \ln x + i 2\pi n} \right\},
\end{equation}
where the remainder term slowly goes to zero as $|n|\to\infty$. 
Let $s={1\over2}\pm{1\over2} \in \{ 0, 1\}$ depending on the sign $\pm$. Then
\begin{eqnarray}
W_n( \pm 2k_0a \; e^{2k_0 a} ) &=& 2k_0 a + \ln(2k_0a) + i (2n+s)\pi 
\nonumber\\
&& + \ln\left[ 2k_0 a + \ln(2k_0a) + i (2n+s)\pi \right] + \dots
\end{eqnarray}
with slowly decreasing remainder  as $|n|\to\infty$. Consequently
\begin{equation}
k_\mathrm{QNF}(n) =  {(2n+s)\pi\over2a} - i { \ln(2k_0 a)\over 2a} - {i\over2a} \ln\left[ 2k_0 a + \ln(2k_0a) + i(2n+s)\pi \right] + \dots 
\end{equation}
Note that it is in this case the real part of the QNF that is equi-spaced as as $|n|\to\infty$, not the imaginary part. Note further that the sub-leading term is logarithmic, not $\mathcal{O}(1/n)$.  This indicates that our intuition regarding the $\hbox{(offset)} + in \hbox{(gap)}$ behaviour built up from black hole mechanics is not quite as universal as one might hope --- at the very least we will have to permit imaginary gaps and offsets, and more general sub-leading terms.


\subsection{Asymmetric double-delta-function potential}
For the asymmetric double delta function
\begin{equation}
V(x) = \alpha_- \,\delta (x - a) + \alpha_+ \, \delta(x + a),
\end{equation}
with the definitions
\begin{equation}
k_+ = {m \alpha_+ \over \hbar^2}; \qquad k_- = {m \alpha_- \over \hbar^2}; 
\end{equation}
we have the transmission amplitude
\begin{equation}
t = {k^2\over (k-ik_+)(k-ik_-) + k_+k_- e^{-4ika} }.
\end{equation}
As $a\to0$ one has
\begin{equation}
t \to  {1\over  1 - i\; \displaystyle{k_++k_-\over k}},
\end{equation}
in agreement with the single delta function case. Also, for $k_+=k_-\to k_0$ this result reproduces the amplitude for the symmetric double-delta potential. 
For real $k$ the transmission probability $T$  can be written as
\begin{eqnarray}  
{k^4 \over k^4 + k^2(k_+^2+k_-^2) + 2 k_+^2k_-^2 + 2k_+k_- [(k^2-k_+k_-)\cos(4ka)  + k(k_++k_-) \sin(4ka)] }.
 \nonumber\\
\end{eqnarray}
This can be recast as
\begin{eqnarray}
T = {1 \over 1 + \displaystyle {(k_+-k_-)^2\over k^2} + {4 k_+k_-\over k^4}  [k \cos(2ka)  +  k_+ \sin(2ka)] \; [k \cos(2ka)  +  k_- \sin(2ka)]  }.
\nonumber\\
\end{eqnarray}
As $a\to0$ one has
\begin{equation}
T \to  {k^2 \over k^2 + (k_++k_-)^2},
\end{equation}
in agreement with the single delta function case. Also, for $k_+=k_-\to k_0$ this result reproduces the amplitude for the symmetric double-delta potential. 

For $k_+\neq k_-$ (and real energies) there are no true transmission resonances as $T$ never quite reaches unity. However \emph{approximate} transmission resonances, where one has a local maximum $T\lesssim 1$, are located at the local minima of the quantity
\begin{eqnarray}
{(k_+-k_-)^2\over 4 k_+ k_-} +  \left[ \cos(2ka)  +  {k_+\over k} \sin(2ka)\right] 
\left[ \cos(2ka)  +  { k_-\over k}  \sin(2ka)\right] \gtrsim 0. 
\end{eqnarray}
For large $k$ the location of these approximate transmission resonances  approximates to
\begin{equation}
\cos(2ka) \approx 0. 
\end{equation}
That is, there is a family of approximate transmission resonances given by 
\begin{equation}
2 ka \approx  \left(n+{1\over2}\right) \pi.
\end{equation}
Near these approximate transmission resonances
\begin{equation}
T \approx {1\over 1 + \displaystyle {(k_+-k_-)^2\over k^2} + {4 k_+^2k_-^2\over k^4} }.
\end{equation}
So these approximate transmission resonances become asymptotically exact as the momentum becomes large.

Turning to complex energies, the QNFs are located by
\begin{equation}
(k-ik_+)(k-ik_-) + k_+k_- e^{-4ika} =0.
\end{equation}
One formal solution is $k=0$, but this does not correspond to a true physical QNF. The non-trivial QNFs do not seem to be explicitly calculable even in terms of the Lambert $W$ function,  but we can develop a perturbative result for  $k_+ \approx k_-$. 

To start the discussion we can at least see that one of the non-trivial QNFs is pure imaginary, (that is, a bound state/purely damped mode). Writing $k=i|k|$ we have
\begin{equation}
(|k|-k_+)(|k|-k_-) = k_+k_- e^{-4|k|a}.
\end{equation}
Graphically, it is easy to see that for $k_+\neq k_-$ there will always be one trivial solution at $k=0$,  plus one distinct non-trivial QNF with $|k|> \max\{k_+,k_-\}$.  

More generally, the exact QNF condition can be rewritten as
\begin{eqnarray}
\left\{ 2ia (k-ik_+)e^{2ika} \right\} \times   \left\{ 2ia (k-ik_-) e^{2ika} \right\} 
= 4 a^2  k_+k_-,
\end{eqnarray}
or
\begin{eqnarray}
\left\{ 2ia (k-ik_+)e^{2i(k-ik_+)a} \right\} \times   \left\{ 2ia (k-ik_-) e^{2i(k-ik_-)a} \right\} 
= 4 a^2  k_+k_- e^{2(k_++k_-) a}.
\qquad
\end{eqnarray}
Defining 
\begin{equation}
C_0 \equiv 2 a \sqrt{k_+ k_-} e^{(k_++k_-) a},
\end{equation}
we have (again in terms of the Lambert $W$ function) 
\begin{equation}
k = i k_+ - i {W_n(C_0 \; e^{+\Theta})\over2a} =  i k_- - i{W_n(C_0 \; e^{-\Theta})\over2a},
\end{equation}
where we have the consistency condition that
\begin{equation}
2a (k_+ - k_-) = W_n(C_0 \; e^{+\Theta} ) - W_n(C_0 \; e^{-\Theta}),
\end{equation}
and for emphasis we have used $W_n$ to indicate that there is an infinite collection of QNFs hiding in the various branches of the Lambert $W$ function. 
For $k_+\approx k_-$ one has a perturbative result
\begin{equation}
a (k_+- k_-) = W_n'(C_0) \Theta + \mathcal{O}(\Theta^3),
\end{equation}
whence
\begin{equation}
 \Theta = {a (k_+- k_-) \over W_n'(C_0)} + \mathcal{O}([k_+-k_-]^3). 
\end{equation}
This easily leads to  the self-consistent estimate
\begin{equation}
k_\mathrm{QNF}(n) =  i \left\{  {k_++k_-\over2} -{ W_n(C_0)\over 2a} + \mathcal{O}([k_+-k_-]^2) \right\} .
\end{equation}
With a bit more work one can extract the next higher-order estimate
\begin{eqnarray}
k_\mathrm{QNF}(n) &=&  i \Bigg\{  {k_++k_-\over2} -{ W_n(C_0)\over 2a} 
-   {a(k_+-k_-)^2\over 4 W_n(C_0) [1+W_n(C_0)]} 
+  \mathcal{O}([k_+-k_-]^4) \Bigg\}.
\nonumber\\
\end{eqnarray}
These results for the perturbative estimation of an infinite tower of QNFs appear to be new. 
The corresponding quasi-normal energies are
\begin{eqnarray}
E_\mathrm{QNF}(n) &=&  - {\hbar^2\over2m} \Bigg\{  {k_++k_-\over2} -{ W_n(C_0)\over 2a} 
-   {a(k_+-k_-)^2\over 4 W_n(C_0) [1+W_n(C_0)]} 
+  \mathcal{O}([k_+-k_-]^4) \Bigg\}^2 .
\nonumber\\
\end{eqnarray}
We could again use the dominant terms of the Comtet expansion, now for $W_n(C_0)$, to determine the asymptotic behaviour as $|n|\to\infty$. We suppress the details as the overall behaviour is qualitatively similar to that for the symmetric double-delta potential considered above.

In contrast, for small $a$ one has a different  perturbative relation (now for the lowest QNF)
\begin{equation}
k_\mathrm{QNF} = i(k_++k_-) -4i k_+k_-a + \mathcal{O}(a^2),
\end{equation}
with 
\begin{equation}
E_\mathrm{QNF} =  - {\hbar^2\over2m} \left\{ (k_++k_-) -4 k_+k_-a + \mathcal{O}(a^2) \right\}^2,
\end{equation}
which is compatible with the result for a single delta-function. 

In summary, all the explicit models we have seen based on one or two delta function potentials lead to a solitary isolated damped mode/bound state (pure imaginary $k_\mathrm{QNF}$), and in addition the situation with two separated delta function potentials leads to an infinite tower of generally complex QNFs.

\subsection{Single-step potential}

The single step potential has the form (see for example Landau--Lifshitz~\cite{Landau}):
\begin{equation}
V(x)  =  \left\{\begin{array}{r@{\quad \quad}l}
V_{0} & (\vphantom{\Big|}\mathrm{for} \quad x>0); \\ 0 & \vphantom{\Big|} (\mathrm{for} \quad x <0).
\end{array} \right.
\end{equation}
Let us write
\begin{equation}
k^2 = {2m  E\over \hbar^2}; \quad\quad q^2 = {2m (E - V_{0}) \over \hbar^2}; 
\quad\quad
k_0^2 =  {2m V_{0} \over \hbar^2} = k^2-q^2.
\end{equation}
Then (see for example Landau--Lifshitz~\cite{Landau}, Messiah~\cite{Messiah}, or Merzbacher~\cite{Merzbacher}, though note different flux conventions) 
\begin{equation}
t = {2  \sqrt{k q}\over k+q}. 
\end{equation}
If $k$ and $q$ are both real (so that $E$ and $E-V_0$ are both real and positive, which is the minimum requirement for a true scattering situation) then the transmission probability is
\begin{equation}
T = {4 k q\over (k+q)^2}  = 1 - {(k-q)^2\over(k+q)^2}. 
\end{equation}
There are no transmission resonances for this potential. (Except formally in the limit $E\to\infty$.)
There are no QNFs for this potential.  From the point of view of QNFs a single-step potential is uninteresting --- but this will change radically once two-step potentials are considered.


\subsection{Rectangular  barrier}
The \emph{rectangular barrier} (sometimes called the \emph{square well}) has the form (see for example Landau--Lifshitz~\cite{Landau}  or Schiff~\cite{Schiff}:
\begin{equation}
V(x)  =  \left\{\begin{array}{r@{\quad \quad}l}
V_{0} & (\vphantom{\Big|}\mathrm{for} \quad  |x| \leq a); \\ 0 & \vphantom{\Big|} (\mathrm{otherwise}).
\end{array} \right.
\end{equation}
Where in general $V_0$ can be either positive or negative.
Let us write
\begin{equation}
k^2 = {2m  E\over \hbar^2}; \quad \quad q^2 = {2m (E - V_{0}) \over \hbar^2}; \quad\quad k_0^2 =  {2m V_{0} \over \hbar^2} = k^2-q^2.
\end{equation}
The exact transmission amplitude is (see for example Brandsen and Joachain~\cite{Brandsen}, or Messiah~\cite{Messiah}, though note change in phase conventions)
\begin{equation}
t = { 4 kq \exp(2ika)  \over (k+q)^2 \exp(2iqa) - (k-q)^2 \exp(-2iqa)   }.
\end{equation}
For real $k$ (real $E$) the exact transmission coefficient is
\begin{equation}
T =  { k^2 q^2 \over  k^2 q^2 + {1\over4} (k^2 - q^2)^2 \, \mathrm{sin}^2 (2qa)},
\end{equation}
which can be rewritten as
\begin{equation}
T = {E (E - V_0) \over E(E - V_0) + {1 \over 4} V_0^{2} \; \mathrm{sin}^2 (2\sqrt{2 m (E - V_0)} a/\hbar)}.
\end{equation}
For more details see (for example) the texts by Landau and Lifshitz~\cite{Landau}, or Schiff~\cite{Schiff}. We can re-write this as
\begin{equation}
T = {1 \over 1 + \displaystyle{ {m V_0^2 a^2 \over 2 E \hbar^2} \, {{\mathrm{sin}^2}(2\sqrt{2 m (E - V_0)} a /\hbar) \over 2 m (E -V_0) a^2/ \hbar^2}} }\, .
\end{equation}
The transmission resonances are defined by
\begin{equation}
T = 1 \qquad \Longleftrightarrow \qquad q = {n\pi\over2a}.
\end{equation}
This now is the first time we see ``families'' of exactly solvable transmission resonances arising. 
In terms of the incident energy the transmission resonances occur at
\begin{equation}
E = V_0 + {\hbar^2 n^2 \pi^2\over 8 m a^2}.
\end{equation}
In contrast the QNFs are located by looking at complex wavenumbers and are defined by $t = \infty$ corresponding to
\begin{equation}
 (k+q)^2 \exp(2iqa) =  (k-q)^2 \exp(-2iqa),
\end{equation}
that is
\begin{equation}
  (k+q) \exp(iqa) =  \pm (k-q) \exp(-iqa),
\end{equation}
which can be simplified to
\begin{equation}
k = - i q \; \tan(qa), \qquad \hbox{or} \qquad k = i q \; \cot(qa), 
\end{equation}
and thence to
\begin{equation}
k_0 = -i q \; \sec(qa), \qquad \hbox{or} \qquad k_0 =  i q \csc(qa),
\end{equation}
whence
\begin{equation}
 q = i k_0  \; \cos(qa), \qquad \hbox{or} \qquad q =- i k_0\; \sin(qa).
\end{equation}
There are two cases of interest for dealing with these transcendental equations.

\subsubsection{Attractive potential} 
For an attractive   potential $V_0<0$, and  then $k_0= i |k_0| $ is imaginary. 
So  for an attractive potential we would be interested in
\begin{equation}
 q = -|k_0|  \; \cos(qa), \qquad \hbox{or} \qquad q = |k_0|\; \sin(qa).
\end{equation}
Apart form the trivial solution at $q=0$, these equations (sometimes) have solutions for real values of $q$, but always with $|q| \leq |k_0|$. Since $k^2 = k_0^2+q^2= -|k_0|^2 + q^2 \leq 0$ this implies pure imaginary values of $k$, which correspond to bound states (not QNFs).

In addition, suppose $Im(q) \gg 0$, then $\cos(qa) \approx \exp(-i q a)/2$ and  $\sin(qa) \approx -\exp(-i q a)/2$. In either case the QNF condition approximates to
\begin{equation}
q \approx - {|k_0|\over 2} \; \exp(-i q a)
\end{equation}
with approximate solution
\begin{equation}
q_\mathrm{QNF}(n) \approx - i {W_n(-i |k_0| a/2) \over a}.
\end{equation}
This approximation should become increasingly accurate for higher values of $Im(q)$, that is, for higher values of $n$.

\subsubsection{Repulsive potential}
On the other hand, for a repulsive potential  $V_0>0$, and  then $k_0>0 $ is real. Let us first consider the  situation $q= i |q|$ (that is, a purely damped mode/ bound state). Then we must solve
\begin{equation}
 |q| =  k_0  \; \cosh(|q|a), \qquad \hbox{or} \qquad q = 0.
\end{equation}
The zero mode is not a true QNF, and the physical QNFs are defined by
\begin{equation}
  |q| a =   k_0 a \; \cosh(|q|a).
\end{equation}
Depending on the precise value of $k_0 a$ there will be two, one, or zero QNFs. For small $k_0 a$ there are two QNFs and one can perturbatively estimate the lower of these two QNFs by 
\begin{equation}
q_\mathrm{QNF} = i k_0 \left\{ 1 + {1\over2}(k_0 a)^2 + {13\over24} (k_0 a)^4 + \mathcal{O}([k_0 a]^6) \right\}.
\end{equation}
This corresponds to
\begin{eqnarray}
k_\mathrm{QNF} = i k_0 (k_0 a)  \left\{ 1+ {2\over3} (k_0 a)^2 + {4\over5} (k_0 a)^4 + \mathcal{O}([k_0 a]^6) \right\}.
\nonumber\\
\end{eqnarray}
The quasi-normal energy is
\begin{eqnarray}
E_\mathrm{QNF} &=&  - {\hbar^2 k_0^2 \over 2m}  (k_0 a)^2  
\left\{ 1 + {4\over3} (k_0 a)^2 +  {92\over45} (k_0 a)^4 + \mathcal{O}([k_0 a]^6) \right\}. \qquad
\end{eqnarray}
In contrast, the higher of these two QNFs seems to have no simple perturbative expansion or other representation.
For $k_0 a \sim 0.663$ the two QNFs merge, (at $qa\sim 1.2$), and for $k_0 a \gtrsim 0.663$ there are no longer any QNFs.

On the other hand, for $Im(q) \gg 0$, we can again approximate the trigonometric functions by exponentials. The QNF condition approximates to
\begin{equation}
q \approx - {i k_0\over 2} \; \exp(-i q a)
\end{equation}
with approximate solution
\begin{equation}
q_\mathrm{QNF}(n) \approx - i {W_n( k_0 a/2) \over a}.
\end{equation}
This approximation should become increasingly accurate for higher values of $Im(q)$, that is, for higher values of $n$.

\subsection{Asymmetric rectangular barrier}
For the \emph{asymmetric rectangular barrier} (sometimes called the \emph{asymmetric  square barrier})  
\begin{equation}
V(x) = \left\{ \begin{array} {r@{, \qquad}l}
V_1 & x<-a; \\ V_{2} & x\in(-a,+a); \\ V_3 & x>a.
\end{array} \right.
\end{equation}
We now define 
\begin{equation}
k_{i} = \sqrt{2 m (E - V_{i})}/\hbar, \qquad i \in\{1,2,3\},
\end{equation}
where the $k_i$ are either pure real or pure imaginary, 
and construct a wave-function of the form (see for instance Messiah, though note change in normalization and phase conventions~\cite{Messiah})
\begin{equation}
\psi(x)  =  \left\{\begin{array}{c@{\quad}l}
\displaystyle{\{\exp(-i k_{1} x) + r \exp(i k_{1} x)\} \over\sqrt{k_1}} & x<-a; \\ 
\\
\displaystyle{\{p \exp(-i k_{2} x) +q  \exp(i k_{2} x)\}\over\sqrt{k_2}} & x\in(-a,+a); \\ 
\\
\displaystyle{t \exp(-i k_{3} x)\over\sqrt{k_3}}  &  x>a.
\end{array} \right.
\end{equation}
The continuity conditions at points $-a$ and $+a$ give the values of $r$, $p$, $q$, and $t$. Without entering into the specific details of the calculation, we simply give the probability amplitude $t$. A brief computation yields:
\begin{equation}
t = {2 k_{2} \sqrt{k_1 k_3} \;  \exp( i [k_{1}+k_3]  a)  \over k_{2} (k_{3} + k_{1}) \mathrm{cos}(2k_{2} a) + i(k_{2}^{2}+ k_{1} k_{3}) \mathrm{sin}(2k_{2}a)}.
\end{equation}
This is more usefully represented as
\begin{equation}
t = {4 k_{2} \sqrt{k_1 k_3} \;  \exp( i [k_{1}+k_3] a)  \over (k_1+k_2)(k_3+k_2)  e^{2i k_{2} a} -  (k_1-k_2)(k_3-k_2)  e^{-2 i k_{2} a} }.
\end{equation}
When $k_1$ and $k_3$ are both real, which is the minimum requirement to have a true scattering problem, the transmission probability is
 (see for example~\cite{Messiah}):
\begin{equation}
T = {4 k_1 k_2^2 k_3 \over (k_1 + k_3)^2 k_2^2 + [k_1^2 k_3^2 + k_2^2(k_2^2 - k_1^2 -k_3^2)] \, \mathrm{sin}^2 (2 k_2 a)} \, .
\end{equation}
The definition of a transmission resonance is now more subtle --- clearly something special happens when the $\sin(\cdot)\to 0$.  Then $T \to T_\mathrm{step}$ --- at this point the transmission probability for the asymmetric rectangular well reduces to that of the transmission probability for a step potential with the same asymptotic behaviour at spatial infinity. 
\begin{equation}
T_{\sin(\cdot)\to 0}  = {4 k_1 k_3 \over (k_1 + k_3)^2} = T_\mathrm{step}(k_1,k_3) .
\end{equation}
(In fact it is known that this step barrier transmission coefficient is a rigorous upper bound on the general transmission probability~\cite{bounds1}.)
We now define the closest we can get to a transmission resonance by asking
\begin{equation}
T/T_\mathrm{step} = 1 \qquad \Longleftrightarrow \qquad 2 k_2 a = n\pi.
\end{equation}
In terms of the energy these ``close as possible to transmission''  resonances occur at
\begin{equation}
E = V_2 + {\hbar^2 n^2 \pi^2\over 8 m a^2}.
\end{equation}
The QNFs are now determined by locating complex wavenumbers for which $t = \infty$, corresponding to 
\begin{equation}
 k_{2} (k_{3} + k_{1}) \mathrm{cos}(2k_{2} a) = - i(k_{2}^{2}+ k_{1} k_{3}) \mathrm{sin}(2k_{2}a).
\end{equation}
That is
\begin{equation}
\tan(2k_2 a) = + i \; {  k_{2} \; (k_{3} + k_{1}) \over (k_{2}^{2}+ k_{1} k_{3}) }.
\end{equation}
Let us now define constants $k_{12}$ and $k_{23}$ (which are either pure real or pure imaginary) by
\begin{eqnarray}
k_{12} =  \sqrt{ {2m(V_2-V_1)\over\hbar^2}}; \qquad k_{23} =  \sqrt{ {2m(V_2-V_3)\over\hbar^2}},\quad
\end{eqnarray}
so that
\begin{eqnarray}
k_1 =  \sqrt{k_2^2 + k_{12}^2}; \qquad k_3 =  \sqrt{k_2^2 + k_{23}^2}.
\end{eqnarray}
Then the QNF condition becomes
\begin{equation}
\tan( 2 k_2 a) = +  i \; {  k_{2} \left(\sqrt{k_2^2 + k_{12}^2} + \sqrt{k_2^2+k_{32}^2} \right) 
\over \left(k_{2}^{2}+ \sqrt{k_2^2 + k_{12}^2} \, \sqrt{k_2^2+k_{32}^2} \right) }.
\end{equation}
This implicitly determines $k_2(a,k_{12},k_{23})$ as a function of the parameters $a$, $k_{12}$, and $k_{23}$. 
This can now be solved perturbatively for $k_2$ as a function of $a$, though the analysis is now considerably more delicate because there are several energy scales in play. 
Provided that both $|k_2| a \ll 1$, (so that we can safely approximate the tangent function by a straight line), and  $|k_{12}^2-k_{23}^2|/|k_{12}^2+k_{23}^2| \ll 1$, (so that the potential is not too asymmetric), one can derive the estimate
\begin{eqnarray}
(k_{2,\mathrm{QNF}})^2 &=& - {1\over4a^2} \; {(k_{12}^2-k_{23}^2)^2\over (k_{12}^2+k_{23}^2)^2} 
- {2 k_{12}^2 k_{23}^2\over k_{12}^2+k_{23}^2} - k_{12}^2 k_{23}^2 a^2 
+ \mathcal{O}(a^4).
\end{eqnarray}
Note that for small $a$ this guarantees a pure imaginary QNF (damped mode/bound state).  
Insofar as the calculations overlap, this agrees with the perturbative QNF estimate for the symmetric rectangular barrier. The corresponding quasi-normal energy is 
\begin{eqnarray}
E_\mathrm{QNF} &=& V_2 - {\hbar^2\over8ma^2} {(V_3-V_1)^2\over(2V_2-V_1-V_3)^2} 
- {2(V_2-V_1)(V_2-V_3)\over (2V_2-V_1-V_3)} 
\nonumber\\
&&
\qquad -  {(V_2-V_1)(V_2-V_3)\over \hbar^2/(2m a^2) }
+ \mathcal{O}(a^4).
\end{eqnarray}
If we now want to move beyond this (small $a$) perturbative  approximation, the algebra unfortunately appears to become intractable. We have not been able to extract anything useful in this case.

\subsection{Tanh potential}
Consider a smoothed step function of the form
\begin{equation}
V(x) =  {V_{-\infty} + V_{+\infty} \over 2} + {V_{+ \infty} - V_{- \infty} \over 2} \; \mathrm{tanh}\bigg({x \over a}\bigg).
\end{equation}
Define
\begin{equation}
k_{\pm\infty} = {\sqrt{ 2m(E-V_{\pm\infty})}\over \hbar};  \qquad \bar k = {k_{+\infty} + k_{-\infty}\over2}; \qquad 
\Delta =  {k_{+\infty} - k_{-\infty}\over2}.
\end{equation}
The transmission amplitude is known analytically to be (see, for instance, \cite{Landau, Birrell-Davies}):
\begin{equation}
t ={\bar k \over \sqrt{k_{+\infty}\; k_{-\infty}  }} \;  {\Gamma(i \bar k a)^2 \over \Gamma(ik_{+\infty} a) \Gamma(ik_{-\infty} a) }.
\end{equation}
Furthermore, the transmission probability is known analytically to be (see for instance \cite{Landau, Birrell-Davies}):
\begin{equation}
T 
= 
{\sinh(\pi  k_{+\infty}a) \sinh(\pi  k_{-\infty}a) \over \sinh^2 (\pi  \bar k a )} 
=
1- {\sinh^2(\pi  \Delta \;\, a) \over \sinh^2(\pi  \bar k\,a)} \, .
\end{equation}
There are no transmission resonances for this potential. (Except formally in the limit $\bar k \to \infty$.) 
To find the QNFs note that  $t = \infty$ when the Gamma function in the numerator has a pole, that is when
\begin{equation}
i  \bar k a= -n,\qquad   n \in \{0,1,2,3,\dots\},
\end{equation}
that is, when
\begin{equation}
i  a (k_{- \infty} + k_{+ \infty})/2 = -n. 
\end{equation}
But we can eliminate one of the asymptotic wavenumbers in terms of the other plus the asymptotic potential difference $V_{+\infty}-V_{-\infty}$. For instance
\begin{equation}
k_{+ \infty} + \sqrt{k_{+ \infty}^{2} + 2 m (V_{+ \infty}-V_{-\infty})/\hbar^2} =  2i  n/ a \, ,
\end{equation}
so that  in terms of the wavenumbers at $x\to+\infty$ the \emph{exact} QNFs are
\begin{equation}
k_{\mathrm{QNF},+ \infty}(n) = i \bigg[{ m (V_{+ \infty}-V_{-\infty}) a \over 2 \hbar^2 \, n} + {n \over a}\bigg]
\qquad n > 0 \, .
\end{equation}
We can completely equivalently specify these \emph{exact} QNFs in terms of the wavenumbers at $x\to-\infty$, in which case
\begin{equation}
k_{\mathrm{QNF},- \infty}(n) = i \,  \bigg[{m  (V_{- \infty}-V_{+\infty})  a\over 2 \hbar^2 \, n} + {n \over a}\bigg] 
\qquad n > 0.
\end{equation}
Though the step from transmission amplitude to QNF is in principle straightforward, an explicit statement as to the exact location of these QNFs seems impossible to find in the extant literature.

Note the asymptotic spacing as $n \rightarrow \infty$:
\begin{equation}
k_{\mathrm{QNF},\pm \infty}(n) \rightarrow i \, {n \over a}.
\end{equation}
Also note that as $a \rightarrow 0$ all the QNFs are driven to imaginary infinity --- this is compatible with the behaviour of the single-step potential for which there are no QNFs.
If one prefers to work with the non-relativistic quasi-normal energies then
\begin{eqnarray}
E_{\mathrm{QNF},n} &=& - {\hbar^2\over 2m a^2} \; n^2 +{ V_{+\infty}+V_{-\infty}     \over 2}
- {m^2a^2(V_{+\infty}-V_{-\infty})^2\over 8 \hbar^2} \; {1\over n^2}.
\end{eqnarray}
Again, though the step from transmission amplitude to this (non-relativistic) quasi-normal energy  is in principle straightforward, an explicit statement as to the exact location of these quasi-normal energies seems impossible to find in the extant literature.

\subsection{Sech$^2$ potential}
Consider a sech$^2$ potential of the form
\begin{equation}
V(x) = V_0 \; \mathrm{sech}^2 (x/a) \, .
\end{equation}
The transmission amplitude is known analytically to be (see for example Landau--Lifshitz~\cite{Landau} or Beyer~\cite{Beyer}):
\begin{eqnarray}
t&=& {
\Gamma\left(i k a + {1\over2} + \sqrt{{1\over4} - 2m V_0 a^2/\hbar^2}\right)
\over  \Gamma(1+ika) 
}
{
\Gamma\left(ik a + {1\over2} - \sqrt{{1\over4} - 2m V_0 a^2/\hbar^2}\right) 
\over \Gamma(ik a)  
}.\quad
\end{eqnarray}
Furthermore, the transmission probability is known analytically to be (see for example Landau--Lifshitz~\cite{Landau}):
\begin{equation}
T = {\mathrm{sinh}^2 [\pi \sqrt{2 m E} a/\hbar] \over \mathrm{sinh}^2[\pi \sqrt{2 m E} a/\hbar] + \mathrm{cos}^2[{1 \over 2} \pi \sqrt{1 - 8 m V_0 a^2/\hbar^2}]}.
\end{equation}
Alternatively
\begin{equation}
T = {\mathrm{sinh}^2 [\pi k a] \over \mathrm{sinh}^2[\pi k a] + \mathrm{cos}^2\left[ \pi \sqrt{{1\over4} - 2 m V_0 a^2/\hbar^2}\right]}.
\end{equation}
Transmission resonances occur but they are not now functions of energy or momentum. Instead $T\to 1$ for
\begin{equation}
\sqrt{1 - 8 m V_0 a^2/\hbar^2} = 2n+1.
\end{equation}
That is, the potential is reflectionless when the depth of the well takes on one of the critical values
\begin{equation}
V_0 = - {\hbar^2  \over 2 m a^2} \; n(n+1).
\end{equation}
The QNFs are in this case quite standard and are presented for instance in Beyer~\cite{Beyer} and many other places in the literature. We have
\begin{equation}
k_{\mathrm{QNF},n} =  i \left\{ {1\over2} \pm \sqrt{{1\over4} - 2m V_0 a^2/\hbar^2} + n \over a \right\} \qquad n\in\{0,1,2,3,\dots\}.
\end{equation}
Note the $\hbox{(offset)}+in\hbox{(gap)}$ behaviour. Note that as $a\to0$ for $V_0$ fixed all but one of the QNFs are driven to imaginary infinity (the one remaining QNF is driven to the unphysical value of zero). In contrast as  $a\to0$ for $V_0 \, a $ held fixed all but one of the QNFs are driven to imaginary infinity and the one remaining QNF is driven to the unique QNF for a delta-function potential.
The corresponding quasi-normal energies are
\begin{equation}
E_{\mathrm{QNF},n} = - {\hbar^2\over2ma^2} \left\{ {1\over2} \pm \sqrt{{1\over4} - 2m V_0 a^2/\hbar^2} + n \right\}^2.
\end{equation}


\subsection[Eckart/ Rosen--Morse/ Morse--Feshbach potential]
{Eckart/ Rosen--Morse/ Morse--Feshbach potential} 

We shall soon see that the Eckart potential can best be viewed as  a linear combination of the (tanh) and  (sech)$^2$ potentials.
To set the stage we emphasize that many of the apparently different potentials commonly encountered in the literature are actually the same quantity in disguise. To start with, consider the following three potentials:

\bigskip
\noindent
{\bf Eckart (1930):}
\begin{equation}
V(x) = - {A \xi \over 1 - \xi} - {B \xi \over (1 - \xi)^2}; \qquad \xi = - \exp(2x/a) .
\end{equation}
{\bf Rosen--Morse (1932):}
\begin{equation}
V(x) = A+ B \, \mathrm{tanh}(x/d) + C \, \mathrm{sech}^2(x/d).
\end{equation}
{\bf Morse--Feshbach (1953):}
\begin{equation}
V(x) = V_0 \; \mathrm{cosh}^2 \mu \; \{\mathrm{tanh}([x - \mu L]/L) + \mathrm{tanh} \, \mu\}^2.
\end{equation}

\bigskip
\noindent
All three of these potentials are actually \emph{identical}.  To see this note that:
\begin{eqnarray}
{4 \xi \over (1 - \xi)^2} &=& {4 \over (\xi^{-1/2}+ \xi^{+1/2})^2} = {4 \over [e^{-x/a} + e^{x/a}]^2}
=
{1 \over \mathrm{cosh}^2(x/a)} = \mathrm{sech}^2(x/a).
\qquad
\qquad
\end{eqnarray}
Similarly
\begin{eqnarray}
1 + {2 \xi \over 1 - \xi} &=& {1 + \xi \over 1 - \xi} = {1 - e^{2x/a} \over 1 + e^{2x/a}} 
={e^{-x/a} - e^{x/a} \over e^{-x/a} + e^{x/a} } = - \mathrm{tanh}(x/a).
\end{eqnarray}
This is enough to show
\[
\hbox{(Eckart)} \Longleftrightarrow \hbox{(Rosen--Morse)}.
\]
In fact in the Rosen--Morse article~\cite{Rosen-Morse}, they cite Eckart~\cite{Eckart}, and describe Eckart's potential as begining ``somewhat like'' their own, but without noticing that the two potentials are in fact \emph{identical} up to trivial redefinitions of the parameters.

Now, for the Morse--Feshbach potential~\cite{Morse-Feshbach}, note that by a trivial shift of origin, $x \rightarrow x + \mu L$, we have 
\begin{equation}
V(x)  \rightarrow V_0 \,  \mathrm{cosh}^2 \mu \{ \mathrm{tanh} (x/L) + \mathrm{tanh} \, \mu \}^2,
\end{equation}
which we can without loss of generality relabel as
\begin{eqnarray}
\nonumber
V(x) &\rightarrow& V_1 \, \{ \mathrm{tanh}(x/L) + D\}^2,
\nonumber
\\
&=& V_1 \{\mathrm{tanh}^2(x/L) + 2 D \, \mathrm{tanh}(x/L) + D^2\},
\nonumber
\\
&=& V_1 \{- \mathrm{sech}^2(x/L) + 2 D \mathrm{tanh}(x/L) + D^2 +1\},
\nonumber
\\
&=& V_2 \,  \mathrm{sech}^2 (x/L) + V_{3} \, \mathrm{tanh} (x/L) + V_4.
\end{eqnarray}
This is enough to show
\[
\hbox{(Morse--Feshbach)} \Longleftrightarrow \hbox{(Rosen--Morse)},
\]
and so all three potentials are completely identical up to trivial relabeling of the parameters and a shift in the zero of energy.

In fact, including the offset, all three of these potentials can be written in \emph{any} one of the four general forms below:
\begin{eqnarray}
V(x) &=& A + B \, \mathrm{tanh} (x/a + \theta) + C \, \mathrm{tanh}^2 (x/a + \theta),
\nonumber
\\
&=& A_0 + [B_0 + C_0 \, \mathrm{tanh} (x/a + \theta)]^2,
\nonumber
\\
&=& A_0 + \bigg[{B_1 + C_1 \, \mathrm{tanh}(x/a) \over B_2 + C_2 \, \mathrm{tanh}(x/a)} \bigg]^2,
\nonumber
\\
&=& A_0 + \bigg[{E_1 + F_1 \, \exp(-2x/a) \over E_2 + F_2 \, \exp(-2 x/a)}\bigg]^2 .
\end{eqnarray}
Note that there is some redundancy here, but it is a useful redundancy. It makes it clear that the Eckart/ Rosen--Morse/  Morse--Feshbach potential is generally the square of a M\"obius function, either of the variable $\tanh(x/a)$ or of the variable  $\exp(-2 x/a)$. Thus implies that without loss of generality we can set either $B_1 C_2 - C_1 B_2$ or $E_1 F_2 - E_2 F_1$ to some convenient constant (often unity).
As long as $E_2$ and $F_2$ do not have opposite signs then the (M\"obius)$^2$ potential does not exhibit any poles, and as long as all of $E_1$, $F_1$, $E_2$, and $F_2$ are in addition nonzero it has the appropriate asymptotic behaviour to define a scattering problem. Other cases are more interesting as model potentials defined on a proper subset of the real line, and possibly with bound states. Sometimes one is interested in formally replacing $a\to ia$ (and taking appropriate real parts) to obtain potentials based in combinations of ordinary trigonometric functions. Sometimes one is interested in formally replacing $a\to -a$, which can also lead to potentials with poles that are defined only on subsets of the real line. For our purposes we will only be interested in those specific forms of the  (M\"obius)$^2$ potential that lead to a well-defined scattering problem.

To explicitly display the transmission amplitude and probability, and the transmission resonances and QNFs, it is convenient to settle on the standard notation
\begin{eqnarray}
V(x) &=&  {V_{-\infty} + V_{+\infty} \over 2} + {V_{+ \infty} - V_{- \infty} \over 2} \; \mathrm{tanh}\bigg({x \over a}\bigg) 
+ {V_0\over\cosh^2(x/a)}.
\\ \nonumber
\end{eqnarray}
That is, the general Eckart potential is simply a linear combination ``(constant) + (tanh) + (sech)$^2$''.
Now define the quantities
\begin{equation}
k_{\pm\infty} = {\sqrt{ 2m(E-V_{\pm\infty})}\over \hbar};  \qquad \bar k = {k_{+\infty} + k_{-\infty}\over2}.
\end{equation}

The transmission amplitude is then known to be (see for example Eckart~\cite{Eckart} or Morse--Feshbach~\cite{Morse-Feshbach})
\begin{eqnarray}
t &=& {-i\over \sqrt{k_{+\infty} \; k_{-\infty}  } \; a} 
\nonumber\\
&& \times  
 {
 \Gamma\left(i \bar k a + {1\over2} + \sqrt{{1\over4} - 2mV_0a^2/\hbar^2} \right)  
\over 
\Gamma(ik_{+\infty} a) 
}
{
 \Gamma\left(i \bar k a  + {1\over2} - \sqrt{{1\over4} - 2mV_0a^2/\hbar^2} \right)  
 \over 
 \Gamma(ik_{-\infty} a) 
 }.
\nonumber\\
\end{eqnarray}
The transmission coefficient is~\cite{Morse-Feshbach}
\begin{equation}
T = { \sinh (\pi k_{- \infty} a) \; \sinh(\pi k_{+ \infty} a) \over 
\sinh^2(\pi \bar k a) + \cos^2\bigg[\pi \sqrt{{1\over4} - {2 m V_0 a^2 \over \hbar^2}}\bigg]}.
\end{equation}
Note that this has appropriate limits as $V_0\to 0$ where it reproduces the $\tanh$ potential, and as $V_{-\infty}\to V_{+\infty} $ where it reproduces the $\sech^2$ potential. 
The closest one now gets to a transmission resonance is that $T\to T_\mathrm{tanh}$ whenever the $\cos[\cdot]\to0$ in the above, that is, whenever the coefficient $V_0$  of the $\sech^2$ part of the potential takes on the special values
\begin{equation}
V_0 = - {\hbar^2  \over 2 m a^2} \; n(n+1).
\end{equation}
Regarding the QNFs, we see that $t\to\infty$ when
\begin{equation}
i \bar k a + {1\over2} \pm \sqrt{{1\over4} - 2m V_0 a^2/\hbar^2} = - n;\qquad \qquad  n\in\{0,1,2,3,\dots\}.
\end{equation}
That is
\begin{equation}
i (k_{- \infty} + k_{+ \infty}) a = \pm  \sqrt{1 - {8 m V_0 a^2 \over \hbar^2}} - (2 n + 1),
\end{equation}
whence
\begin{equation}
(k_{- \infty} + k_{+ \infty}) = \pm i {1 \over a} \sqrt{1 - {8 m V_0 a^2 \over \hbar^2}} + i {2 n + 1 \over a}.
\end{equation}
We can now rearrange this solely in terms of $k_{+\infty}$ by writing
\begin{eqnarray}
k_{+ \infty} + \sqrt{k_{+ \infty}^2 + 2m (V_{+ \infty}-V_{-\infty})/\hbar^2} 
= \pm i {1 \over a} \sqrt{1 - {8 m V_0 a^2 \over \hbar^2}} + i {2 n + 1 \over a} ,
\end{eqnarray}
implying
\begin{eqnarray}
k_{+ \infty} + \sqrt{k_{+ \infty}^2 + 2m (V_{+ \infty}-V_{-\infty})/\hbar^2} 
= i \left({2 n +1 \pm \sqrt{1 - 8 m V_0 a^2/\hbar^2} \over a}\right).
\qquad
\end{eqnarray}
Note that this has appropriate limits for the tanh and sech$^2$ potentials. Finally, because this is a simple quadratic equation, we can solve for the QNFs $k(n)$ to obtain the \emph{exact} QNFs
\begin{eqnarray}
k_{\mathrm{QNF},+\infty}(n) &=& i\Bigg(  {m (V_{+\infty}-V_{-\infty}) a / \hbar^2 \over (2n + 1) 
\pm \sqrt{1 - 8 m V_0 a^2 /\hbar^2}} 
+  {(2 n + 1) \pm \sqrt{1 - 8 m V_0 a^2/ \hbar^2} \over 2 a}\Bigg).
\nonumber\\
\end{eqnarray}
A similar analysis in terms of the wavenumber  $k_{-\infty}$ at $x\to-\infty$ casts the \emph{exact} QNFs in the form
\begin{eqnarray}
k_{\mathrm{QNF},-\infty}(n) &=& i\Bigg({ m (V_{-\infty}-V_{+\infty}) a/\hbar^2 \over (2 n+1) \pm \sqrt{1 - 8 m V_0 a^2/ \hbar^2} } 
+ {(2 n+1) \pm \sqrt{1 - 8 m V_0 a^2/ \hbar^2} \over 2 a}\Bigg).
\nonumber\\
\end{eqnarray}
This now has the appropriate limits to reproduce both tanh and sech$^2$ QNFs. We have not found explicit formulae of this type in the extant literature. 
Note that asymptotically
\begin{equation}
k_{\mathrm{QNF},\pm\infty}(n) \rightarrow i \, {n \over a} + i \, {1 \pm \sqrt{1 - 8 m V_0 a^2/ \hbar^2}  \over 2 a} + \mathcal{O}(1/n),
\end{equation}
in accordance with the general suspicions based on black hole QNMs~\cite{Medved1, Medved2, Padmanabhan1, Padmanabhan2, Jozef1, Jozef2, Jozef3, Jozef4}. Finally we consider the quasi-normal energy
\begin{eqnarray}
E_\mathrm{QNF}(n) &=& V_{+\infty} + {\hbar^2 k_{\mathrm{QNF},+\infty}(n)^2  \over 2m} =
V_{-\infty} + {\hbar^2 k_{\mathrm{QNF},-\infty}(n)^2  \over 2m},
\end{eqnarray}
and compute
\begin{eqnarray}
E_\mathrm{QNF}(n) &=& -  { m a^2\over2\hbar^2} \; { (V_{+\infty}-V_{-\infty})^2 \over \left( 2 n +1 \pm \sqrt{1 - 8 m V_0 a^2/ \hbar^2}\right)^2}
\nonumber\\
&& + {V_{+\infty}+V_{-\infty}\over2} 
-  {\hbar^2\over 8ma^2}  \left(2 n +1 \pm \sqrt{1 - 8 m V_0 a^2/ \hbar^2}\right)^2.
\end{eqnarray}
Again, we have not found explicit formulae of this type in the extant literature.

\subsection[Related potentials\\  (Morse, P\"oschl--Teller, Manning--Rosen, Hulthen, Teitz, Hua)]
{Related potentials \\ (Morse, P\"oschl--Teller, Manning--Rosen, Hulthen, Teitz, Hua) }
A number of  other closely related  potentials are also of some interest --- though most often one is simply revisiting the Eckart potential in disguise. 

\subsubsection{Morse (1929)} 
Consider the potential
\begin{equation}
V(x) = V_0 \, (1 - \exp (- [x - x_0]/a))^2.
\end{equation}
This  Morse potential is actually a somewhat odd limit of the (M\"obius)$^2$ potential as various parameters go to unity or zero.
It is most useful as a model for bound states and does not define a scattering problem.

\subsubsection{P\"oeschl--Teller (1933)}
We should first warn the reader that the actual article by P\"oschl and Teller~\cite{Poeschl-Teller}  is somewhat difficult to get hold of.  That article starts by discussing the potential
\begin{equation}
V(x) = V_0 \left\{  {A\over\sin^2(x/a)} + {B\over\cos^2(x/a) }\right\}; \qquad x \in (0, \pi a/2),
\end{equation}
and its hyperbolic analytic continuation ($a\to ia$) 
\begin{equation}
V(x) = V_0 \left\{  {A\over\sinh^2(x/a)} + {B\over\cosh^2(x/a) }\right\}; \qquad x \in (0, \infty),
\end{equation}
relying only on a change of \emph{font} (typeface) to make the distinction between ordinary and hyperbolic trigonometric functions.  Finally the article focusses  attention on the specific case
\begin{equation}
V(x) = {V_0   \over\cosh^2(x/a) }   = V_0 \; \mathrm{sech}^2 (x/a); \qquad x \in (-\infty, \infty). 
\end{equation}
Because of this many authors use the phrase ``P\"oschl--Teller potential'' to refer to the $\sech^2$ potential. 
While this is historically somewhat inaccurate, insofar as the $\sech^2$ potential is already contained as a special case of the Eckart potential, this terminology now seems firmly embedded in the literature. (Oddly, P\"oschl and Teller refer to both the Morse 1929~\cite{Morse} and Rosen--Morse 1932~\cite{Rosen-Morse} articles, but not the Eckart 1930~\cite{Eckart} article.) 

\subsubsection{Manning--Rosen (1933)}
Consider the potential
\begin{equation}
V(x) = B \, \mathrm{coth}(x/a) - C \, \mathrm{cosech}^2 (x/a); \qquad  x\in(0, \infty).
\end{equation}
The relevant citation~\cite{Manning} is only an \emph{abstract} in a report of a conference. To find it with online tools such as {\sf PROLA} look up Phys.~Rev.~{\bf 44}\, (1933) \,951, and then manually scan for abstract $\# \, 10$.  The form actually given in the abstract is
\begin{equation}
V(x) = A {\exp(-2 x/b) \over [1 - \exp(-x/b)]^2} + B \, {\exp(-x/b) \over 1 - \exp(-x/b)},
\end{equation}
which one can manipulate into the form above by noting
\begin{eqnarray}
1 + 2 \, {\exp(-x/b) \over 1 - \exp(-x/b)} &=& {1 + \exp(-x/b) \over 1 - \exp(-x/b)}
={\exp(x/2b) + \exp(-x/2b) \over \exp(x/2b) - \exp(-x/2b)} 
=\mathrm{coth}(x/2b),
\nonumber\\
\end{eqnarray}
and 
\begin{equation}
\mathrm{coth}^2 z = 1 + \mathrm{cosech}^2 z.
\end{equation}
Note that the Manning--Rosen potential can be obtained from the Eckart potential by the formal substitution $x \to -x+ i\pi a /2$ so that
\begin{equation}
\xi = - \exp(2x/a) \rightarrow + \exp(-2x/a).
\end{equation}
In particular, Manning--Rosen can be written in the form
\begin{eqnarray}
V(x) &=& A + B \, \mathrm{coth} (x/a) + C \, \mathrm{coth}^2 (x/a)
=A_0 + [B_0 + C_0 \, \mathrm{coth}(x/a)]^2.
\qquad
\end{eqnarray}
We can get this from the general (M\"obius)$^2$ form of the Eckart potential by appropriately choosing the parameters. Because of the pole at $x=0$ the potential is best thought  of as being defined on $(0,\infty)$. It does not define a scattering problem, though it may be useful for investigating bound states. 

\subsubsection{Hulthen (1942)}
Consider the potential
\begin{equation}
V(x) = V_0 \, {\exp(-x/a) \over 1 - \exp(-x/a)}; \qquad  x\in(0, \infty).
\end{equation}
This  Hulthen potential~\cite{Hulthen} is actually a special case of the Manning--Rosen potential. We can also get this from the general (M\"obius)$^2$ form of the Eckart potential by appropriately choosing the parameters. Because of the pole at $x=0$ the potential is best thought  of as being defined on $(0,\infty)$. It does not define a scattering problem, though it may be useful for investigating bound states. 

\subsubsection{Tietz (1963)}
One version of the Tietz potential~\cite{Tietz} is:
\begin{equation}
V(x) = V_0 \left({\mathrm{sinh}([x - x_0]/a) \over \{\mathrm{sinh, cosh, \exp}\} (x/a)}\right)^2.
\end{equation}
We can get this from the general (M\"obius)$^2$ form of the Eckart potential by appropriately choosing the parameters.
Depending on the specific choices made it may or may  not define a scattering problem, though it may be useful for investigating bound states. 

\subsubsection{Hua (1990)}
Hua's potential is~\cite{Hua}
\begin{equation}
V(x) = V_0 \bigg({1 - \exp(-2x/a) \over 1 - q \, \exp(-2 x/a)}\bigg)^2.
\end{equation}
We can get this~\cite{Hua, Natanson}  from the general (M\"obius)$^2$ form of the Eckart potential by appropriately choosing the parameters. We note
\begin{eqnarray}
V(x) &=& V_0 \bigg({\exp(x/a) - \exp(-x/a) \over \exp(x/a) - q \, \exp(-x/a)}\bigg)^2,
\nonumber
\\
&=& V_0 \bigg({\mathrm{sinh} (x/a) \over (1+q) \mathrm{sinh}(x/a) + (1-q) \mathrm{cosh}(x/a)}\bigg)^2.
\nonumber\\
\end{eqnarray}
If $q > 0$ define $(1- q)/(1+q) = \tanh \theta$.  If $q < 0$ define $(1+ q)/(1-q) = \tanh \theta$. 
Then we see
\begin{eqnarray}
V(x) &=& V_1 \bigg({\mathrm{sinh}(x/a) \over \{\mathrm{sinh, cosh}\}(x/a + \theta)}\bigg)^2 
\qquad (q \neq 0) ,
\nonumber
\\
&=& V_1 \bigg({\mathrm{sinh}(\bar{x}/a - \theta) \over \{\mathrm{sinh, cosh, \exp}\} (\bar{x}/a)}\bigg) ^2,
\nonumber
\\
&=& \hbox{(Tietz potential)},
\nonumber
\\
&=& V_1 \bigg({A \, \mathrm{sinh}(\bar{x}/a) + B \, \mathrm{cosh} (\bar{x}/a) \over \{\mathrm{sinh, cosh}\}(\bar{x}/a)}\bigg)^2,
\nonumber
\\
&=& (\hbox{Eckart/ Manning--Rosen  as appropriate}).
\nonumber\\
\end{eqnarray}

\paragraph{Summary:} So all of these potentials are either identical to the (M\"obius)$^2$ potential, or special cases of the (M\"obius)$^2$ potential. To be historically accurate we should really just call this whole collection of potentials the Eckart potential, or appropriate special cases of the Eckart potential, as Eckart seems to have been the first author to have given the general form. Unfortunately other names are now in such common use that historical accuracy is difficult (if not impossible) to recover.

\begin{table*}[!ht]
\label{T:comparisons}
\centerline{Inter-relationships between various ``exactly solvable'' potentials}
\bigskip
\hskip-1cm
{\small
\begin{tabular}{|  l  |  l  |  l |}
\hline
Name & Potential $V(x)$ & Properties \\
\hline
\hline
\vphantom{\bigg|} Morse $(1929)$ & $V_0 \, (1 - \exp (- x/a))^2$   & 
Special limit of Eckart/(M\"obius)$^2$ \\
\hline
\hline
\vphantom{\Bigg|}  Eckart $(1930)$ & $\displaystyle{ - {A\, \xi \over 1 - \xi} - {B\, \xi \over (1 - \xi)^2} }$;   \quad $\xi = - \exp(2x/a)$ & 
$
\Leftrightarrow  \hbox{Rosen--Morse} \Leftrightarrow \hbox{(M\"obius)}^2$ \\
\hline
\vphantom{\bigg|} Rosen--Morse $(1932)$ & $A+ B \, \mathrm{tanh}(x/a) + C \, \mathrm{sech}^2(x/a)$ & 
$ 
\Leftrightarrow \hbox{Eckart}  \Leftrightarrow \hbox{(M\"obius)}^2$ \\
\hline
\vphantom{\bigg|}  Morse--Feshbach $(1954)$ & $V_0 \, \mathrm{cosh}^2 \mu  \{\mathrm{tanh}([x - \mu a]/a) + \mathrm{tanh} \, \mu\}^2 $ & 
$  
\Leftrightarrow \hbox{Rosen--Morse}  \Leftrightarrow \hbox{Eckart}$ \\
\hline
\vphantom{\bigg|}  Eckart/ Rosen--Morse & $A + B \, \mathrm{tanh} (x/a ) + C \, \mathrm{tanh}^2 (x/a)$ 
& $\Leftrightarrow$ (M\"obius)$^2$  function of   $\exp(-2 x/a)$ \\
\hline
\vphantom{\Bigg|} (M\"obius)$^2$ & $  \displaystyle{ V_0 \bigg[{A + B \, \exp(-2x/a) \over C + D \, \exp(-2x/a)}\bigg]^2}  $ 
& The ``best'' of these equivalent forms \\
\hline
\hline
\vphantom{\bigg|} Manning--Rosen $(1933)$ & $A+ B \, \mathrm{coth}(x/a) - C \, \mathrm{cosech}^2 (x/a)$ & Special limit of Eckart/(M\"obius)$^2$ \\
\hline
\vphantom{\Bigg|} Hulthen $(1942)$ & $\displaystyle{ V_0 \, {\exp(-x/a) \over 1 - \exp(-x/a)}} $ & Special case of Manning--Rosen \\
\hline
\vphantom{\Bigg|} Tietz $(1963)$ &  $\displaystyle{ V_0 \bigg({\mathrm{sinh}([x - x_0]/a) \over \{\mathrm{sinh, cosh, \exp}\} (x/a)}\bigg)^2}$ 
& Special limit of Eckart/(M\"obius)$^2$ \\
\hline
\vphantom{\Bigg|} Hua $(1990)$ & $\displaystyle{ V_0 \bigg({1 - \exp(-2x/a) \over 1 - q \, \exp(-2 x/a)}\bigg)^2} $  
& $\hbox{Eckart  or  Manning--Rosen or Morse}$ \\
\hline

\hline
\end{tabular}
} 
\caption[Inter-relationships between selected ``exactly solvable'' potentials]{\label{T:compare-4} This table shows the inter-connections between many ``exactly solvable'' potentials. Many of these potentials are \emph{identical} to each other, though this may not always be obvious at first glance.} 
\bigskip
\end{table*}

\section{Discussion}

In this paper, we have collected many known analytic results, and described several significant new results on analytic QNFs, in a form amenable to comparison with the extant literature. In particular we have, in addition to the QNFs themselves, focussed on transmission amplitudes, transmission probabilities, and transmission resonances.
We did this explicitly for the delta--function potential, double--delta--function potential, and asymmetric double--delta--function potential; the  step barrier, rectangular  barrier, and asymmetric rectangular barrier; the tanh potential, sech$^2$ potential,  and Eckart potential and its variants. In almost all of these cases we have been able to take the calculation of the QNFs somewhat further, sometimes significantly further,  than currently available sources.

In particular, we have noted that the Eckart/Rosen--Morse/Morse-Feshbach  potentials are actually \emph{identical}, and that they are generally a (M\"obius)$^2$ function of the variable  $\exp(-2 x/a)$.  Indeed many of the ``exactly solvable'' potentials commonly encountered in the literature are actually the same quantity in disguise, typically the (M\"obius)$^2$ potential itself or some special case thereof --- and so really should just be collectively referred to as variants of the Eckart potential. 
We should also mention that there has recently been some progress in analyzing the (approximate highly damped) QNFs for \emph{piecewise} Eckart potentials~\cite{Jozef1, Jozef2, Jozef3, Jozef4}. 

What message can we extract concerning the commonly conjectured ``$\hbox{(offset)} + i n \hbox{(gap)}$'' behaviour for highly damped QNFs? From the examples we have seen here (and in~\cite{Jozef1, Jozef2, Jozef3, Jozef4}) is appears that the ``$\hbox{(offset)} + i n \hbox{(gap)}$'' behaviour depends on both a non-zero width for the potential, and a certain amount of smoothness. The double-delta potential leads to imaginary gap, imaginary offset, and logarithmic sub-leading terms --- which is not what one would naively have expected. 

Finally, we reiterate that very few potentials have exact analytically known quasi-normal frequencies [QNFs]. Even for so-called ``analytically solvable'' potentials  it is not necessarily true that the QNFs can be explicitly located in closed form. Thus apart from their intrinsic interest, these exact and approximate results serve as a backdrop and a consistency check on ongoing efforts to locate and understand QNFs in general physical situations.

\acknowledgments

This research was supported by the Marsden Fund administered by the Royal Society of New Zealand. PB was additionally supported by a scholarship from the Royal Government of Thailand, and partially supported by a travel grant from FQXi, and by a grant for the professional development of new academic staff from the Ratchadapisek Somphot Fund at Chulalongkorn University. 


\end{document}